\documentclass{svjour2}                    
\smartqed  
\usepackage{amsmath,amssymb}
\usepackage{graphicx}
\begin{document}

\title{On the principle of minimum growth rate in multiplicatively interacting stochastic processes}
\titlerunning{Principle of min. growth rate in multiplicatively interacting proc.}        

\author{Akihiro Fujihara \and Toshiya Ohtsuki  \and Hiroshi Yamamoto}

\authorrunning{A. Fujihara, T. Ohtsuki, H. Yamamoto} 
\institute{A. Fujihara$^*$, T. Ohtsuki, and H. Yamamoto \at
              Field of Natural Sciences, International Graduate College of Arts and Sciences, Yokohama City University, 22-2 Seto, Kanazawa-ku, Yokohama 236-0027, Japan \\
              \email{$^*$fujihara@yokohama-cu.ac.jp} 
}

\date{Received: date / Accepted: date}

\maketitle

\begin{abstract}
A method of moment inequalities is used to derive the principle of minimum growth rate in multiplicatively interacting stochastic processes(MISPs). When a value of a power-law exponent at the tail of probability distribution function exists in a range $0 < s \le 1$, a first-order moment diverges and an equality for a growth rate of systems breaks down. From the estimate of inequalities, we newly find a conditional inequality which determines the growth rate, and then the exponent in $0 < s \le 1$. 
\end{abstract}

\section{Introduction}
\label{intro}
 Multiplicatively interacting stochastic processes(MISPs) have been investigated quite actively for past some years\cite{BCG2000,EB2002,BK2002,BMP2002,BBLR2003,PB2003,IKR1998,Slanina2004,BGP2004,OFY2004,FOY2004,CC2005,FTOY2005}. It is known that these processes generate power-law distributions at the tail of a probability distribution function(PDF). They are considered one of probable candidates for explaining power-law tails observed widely in natural phenomena\cite{Sornette2004,Newman2005}, such as velocity distributions of inelastically colliding particles and wealth distributions in economics. There exist two parameters characterizing the processes: a growth rate of systems and a power-law exponent. These are able to be calculated analytically. It is found that both parameters vary continuously depending on microscopic interaction parameters in a certain region. This fact shows that the power-law behavior of MISPs is vastly different from that of usual critical phenomena. In critical phenomena, power-law behavior appears when the parameters are controlled only at a critical point. In MISPs, on the other hand, a power-law tail emerges in a certain region of the parameter space, which suggests ubiquity of power laws due to MISPs. The second difference is universality. In the former, the power-law exponent is independent of microscopic parameters, and therefore becomes an indicator of universality classes. In the latter, the exponent depends on microscopic details and universality does not hold. It is known that one-body random multiplicative models also show power-law tails in some conditions\cite{Sornette2004,TST1997}. The dependence of the exponent on interaction parameters indicates that many-body processes is irreducible to one-body processes because interaction parameters are specific for many-body ones. It will be reported that in weighted interactions, many-body processes belong to a different universality class from that of one-body ones\cite{FTOY2005}. 

The paper is organized as follows. In Sec.~\ref{model}, we introduce a method of Fourier transform which is usually used for calculating the growth rate $\gamma$ and the exponent $s$ of granular gas systems\cite{BCG2000,EB2002,BK2002,PB2003} and other MISPs\cite{BBLR2003,Slanina2004,OFY2004,FOY2004}. In Sec.~\ref{principle}, we explain the principle of minimum growth rate which is originally found by ben-Avraham \textit{et al.}\cite{BBLR2003}. In Sec.~\ref{moment-ineq}, we apply a method of moment inequalities which is used in inelastic hard-sphere models\cite{BGP2004} in order to derive the principle theoretically. In Sec~\ref{summary-discussion}, summary and discussions are stated.

\section{Model and method of Fourier transform}
\label{model}
\begin{figure}
\resizebox{60mm}{!}{\includegraphics{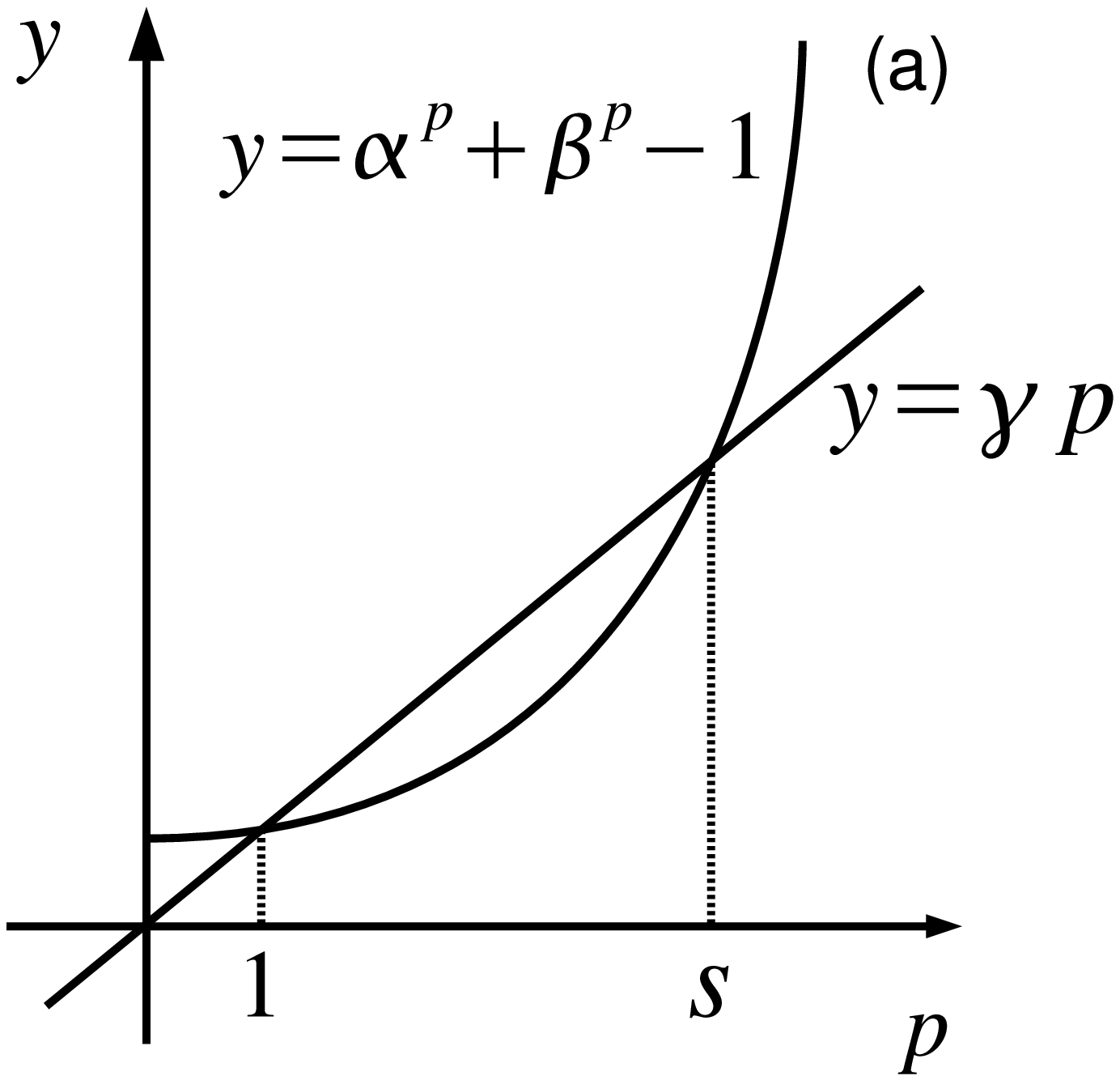}}
\resizebox{60mm}{!}{\includegraphics{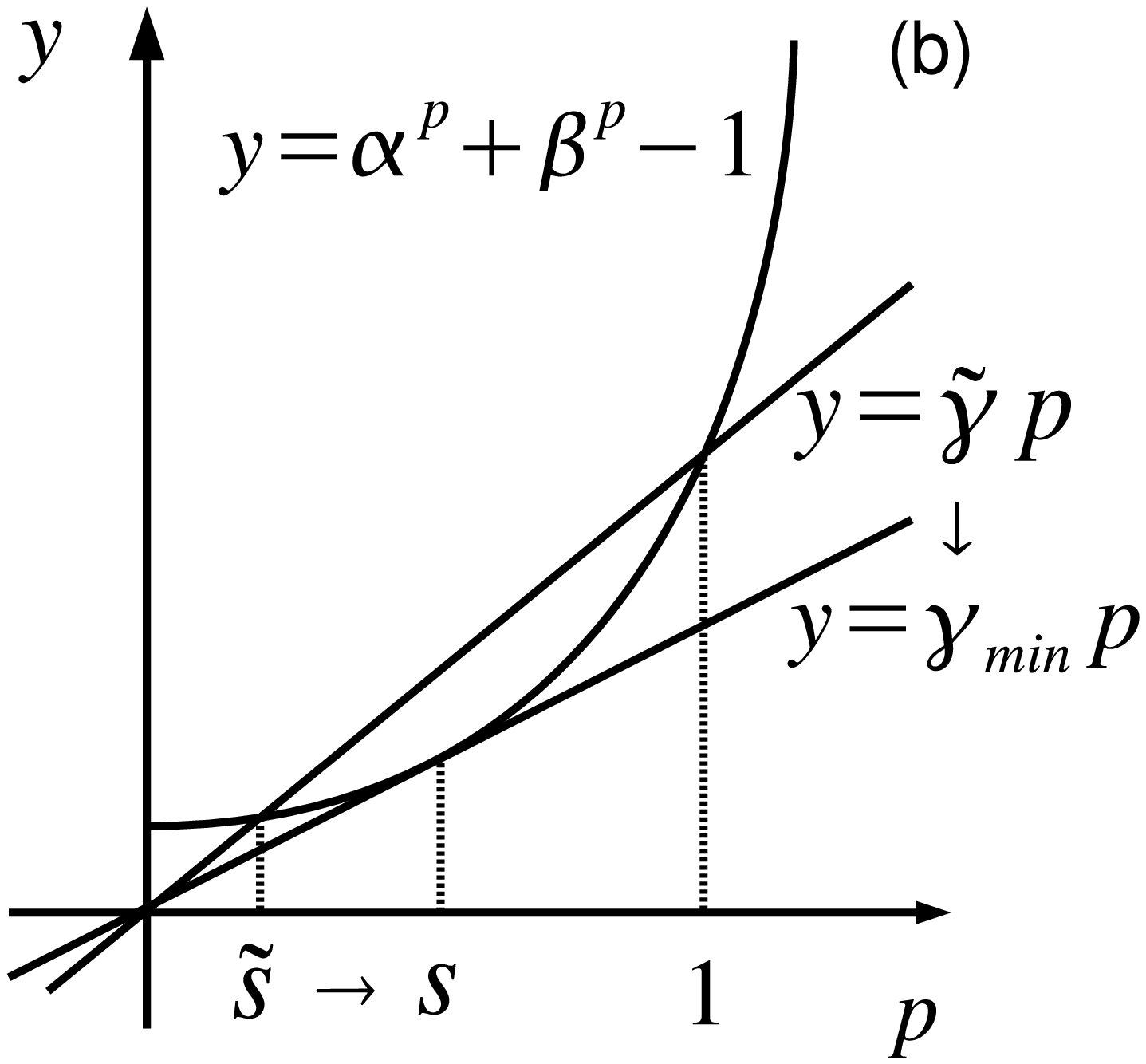}}
\caption{Solution of the transcendental equation at $s > 1$ (a) and at $0 < s \le 1$ (b). In case (b), the principle of minimum growth rate shows that the growth rate is given by not $\tilde{\gamma}$ but $\gamma_{min}$, and correspondingly the power-law exponent is not $\tilde{s}$ but $s$.} 
\label{fig:minselect}
\end{figure}

 Here, let us introduce the method of Fourier transform to derive the growth rate and power-law exponent\cite{BCG2000,EB2002,BK2002,BBLR2003}. We consider a $N-$particle system defined as follows. Each particle has positive quantities $x_{i} > 0$ ($i = 1, \cdots, N$). At each time step, two particles labeled by $i$ and $j$ ($i \neq j$) are picked up randomly from $N$ particles and interact with each other, where two quantities $x_{i}, x_{j}$ are transformed into $x'_{i}, x'_{j}$ by the rule 
\begin{eqnarray}
x_{i}' = \alpha x_{i} + \beta x_{j}, \hspace{5mm} x_{j}' = \beta x_{i} + \alpha x_{j},
\end{eqnarray}
with positive interaction parameters $\alpha, \beta (>0)$. When $N \to \infty$, this process is described by the following master equation. 
\begin{eqnarray}
\frac{\partial f(z,t)}{\partial t} &=& \int_{0}^{\infty}dx \int_{0}^{\infty}dy \ f(x,t) \ f(y,t) \\
				   & & \times \frac{1}{2} \left[\right. \delta( z - ( \alpha x + \beta y ) ) + \delta( z - ( \beta x + \alpha y ) ) - \delta( z - x ) - \delta( z - y ) \left.\right], \nonumber \label{eq:mastereq-z} 
\end{eqnarray}
where $f(z, t)$ is a PDF of quantities. After the Fourier transform $F(k, t) = \int_{0}^{\infty} dz e^{ikz} f(z, t)$, Eq.~(\ref{eq:mastereq-z}) becomes 
\begin{eqnarray}
\frac{\partial F(k,t)}{\partial t} = F(\alpha k, t) F(\beta k, t) - F(k, t). \label{eq:mastereq-k}
\end{eqnarray}
 In this model, the total value of the quantity is not conserved. In the long time limit, therefore, the PDF diverges to infinity or shrinks to zero depending on the value of interaction parameters $\alpha, \beta$. Then, the scaling relations 
\begin{eqnarray}
\xi = z e^{ - \gamma t }, \hspace{5mm} \psi( \xi ) = f(z, t) e^{ \gamma t }, \label{eq:scaling-relation-z} \\
\kappa = k e^{\gamma t}, \hspace{5mm} \Psi( \kappa ) = F(k, t), \label{eq:scaling-relation-k}
\end{eqnarray}
are assumed, where $\gamma$ is a scaling parameter representing a growth rate of systems. Substituting Eq.~(\ref{eq:scaling-relation-k}) into Eq.~(\ref{eq:mastereq-k}), we obtain 
\begin{eqnarray}
\gamma \kappa \frac{d \Psi(\kappa)}{d \kappa} = \Psi(\alpha \kappa) \Psi(\beta \kappa) - \Psi(\kappa). \label{eq:mastereq-k-scaled}
\end{eqnarray}
 Differentiating Eq.~(\ref{eq:mastereq-k-scaled}) with respect to $\kappa$, and then substituting $\kappa = 0$, we obtain the growth rate of systems 
\begin{eqnarray}
\gamma = \alpha + \beta - 1. \label{eq:gamma}
\end{eqnarray}
 Performing the same calculations in Eq.~(\ref{eq:mastereq-k}), we find $\gamma$ is actually the growth rate of the first-order moment, that is, $\int_{0}^{\infty} z f(z, t) dz \sim e^{\gamma t}$. Suppose $\Psi(\kappa)$ consists of a combination of regular and singular terms, 
\begin{eqnarray}
\Psi(\kappa) &=& \Psi_{regular}(\kappa) + \Psi_{singular}(\kappa), \\
	     &=& \sum_{n=0}^{\infty} \frac{(i \kappa)^{n}}{n!}a_{n} + a_{p} \kappa^{p}, 
\end{eqnarray}
 where $p$ is a non-integer number in general. In this case, the small leading term of $\Psi_{singular}(\kappa)$ reflects the power-law tail $\psi(\xi) \simeq 1/\xi^{1+s}$, where $s$ is a power-law exponent. Substituting $\Psi(\kappa) \simeq 1 + a_{p} \kappa^{p}$ into Eq.~(\ref{eq:mastereq-k-scaled}), we find a transcendental equation 
\begin{eqnarray}
p \gamma = \alpha^{p} + \beta^{p} - 1. \label{eq:teq}
\end{eqnarray}
 From this transcendental equation together with Eq.~(\ref{eq:gamma}), we finally obtain a non-trivial solution $p = s (\neq 1)$ as illustrated in Fig.~\ref{fig:minselect}(a). 
 This solution corresponds to the power-law exponent. In this way, the growth rate of systems $\gamma$ and the power-law exponent $s$ are calculated analytically.

\section{Principle of minimum growth rate}
\label{principle}
 In the case $0 < s \le 1$, the method of Fourier transform mentioned above does not work because the first-order moment diverges, and thus Eq.~(\ref{eq:gamma}) does not hold anymore. Just one equation (\ref{eq:teq}) remains to hold and two parameters $\gamma$ and $s$ becomes indeterminate. ben-Avraham \textit{et al.}\cite{BBLR2003} discovered that when $0 < s \le 1$, $\gamma$ and $s$ are determined to fulfill the following relations.
\begin{eqnarray}
s \gamma_{min} = \alpha^{s} + \beta^{s} - 1, \hspace{5mm} \gamma = \gamma_{min} = \alpha^{s} \ln \alpha + \beta^{s} \ln \beta. \label{eq:minselect-teq}
\end{eqnarray}
 These relations indicates that the value of the growth rate $\gamma$ is equal to the tangential gradient of $y = \alpha^{p} + \beta^{p} - 1$ going through the origin as shown in Fig.~\ref{fig:minselect}(b). In other words, it can be said that the growth rate $\gamma$ becomes minimum under the condition that Eq.~(\ref{eq:teq}) has a solution. For this reason, we call this mechanism \textit{the principle of minimum growth rate}. To our knowledge, the principle has never been explained theoretically. In this paper, we derive the principle by applying a method of moment inequalities, instead of that of Fourier transform.

\section{Method of moment inequalities}
\label{moment-ineq}
 Substituting Eq.~(\ref{eq:scaling-relation-z}) into Eq.~(\ref{eq:mastereq-z}), we obtain 
\begin{eqnarray}
p \gamma \mu_{p} = \int_{0}^{\infty} d\xi_{1} \int_{0}^{\infty} d\xi_{2} \psi(\xi_{1}) \psi(\xi_{2})\left[\right. ( \alpha \xi_{1} + \beta \xi_{2} )^{p} - (\xi_{1})^{p} \left.\right]. \label{eq:rescaled-mastereq}
\end{eqnarray}
 where $\mu_{p}$ is the $p-$th order moment of a scaled PDF $\psi(\xi)$. The growth rate of systems Eq.~(\ref{eq:gamma}) is derived from Eq.~(\ref{eq:rescaled-mastereq}) at $p=1$. Substituting the inequalities of lemma $2$ in the paper by Bobylev \textit{et al.}\cite{BGP2004}, we find the inequalities of moments 
\begin{eqnarray}
\frac{ A(\alpha, \beta, p) }{ \gamma p - ( \alpha^{p} + \beta^{p} - 1) } \le \mu_{p} \le \frac{ B(\alpha, \beta, p) }{ \gamma p - ( \alpha^{p} + \beta^{p} - 1) }. \label{eq:moment-inequalities}
\end{eqnarray}
 for all $p > 0$, where $A, B$ are certain bounded functions. Equation (\ref{eq:moment-inequalities}) means that the solution $p=s$ of Eq.~(\ref{eq:teq}) represents the order of divergent moment $\mu_{s}$, and thus the power-law exponent $s$. In this way, $\gamma$ and $s$ are derived from Eqs.~(\ref{eq:gamma}) and (\ref{eq:teq}) in the case $s > 1$. Therefore, this method leads to the same results as the method of Fourier transform did. In the case $0 < s \le 1$, the next inequality is satisfied. 
\begin{eqnarray}
(x + y)^{p} \le x^{p} + y^{p}, \label{eq:xy-ineq}
\end{eqnarray}
 for all $0 < p \le 1$, where $x, y > 0$. Substituting Eq.~(\ref{eq:xy-ineq}) into Eq.~(\ref{eq:rescaled-mastereq}), we obtain the conditional inequality. 
\begin{eqnarray}
p \gamma \le \alpha^{p} + \beta^{p} - 1, \label{eq:gamma-condition}
\end{eqnarray}
 for all $0 < p \le 1$. It follows that when the power-law exponent exists in a region $0 < s \le 1$, the growth rate $\gamma$ and the exponent $s$ is determined so as to fulfill the next two conditions. 
\vspace{2mm}
\begin{enumerate}
\item There exist a solution $p=s$ in $0 < p \le 1$ satisfying $p \gamma = \alpha^{p} + \beta^{p} - 1$.
\vspace{1mm}
\item Condition $p \gamma \le \alpha^{p} + \beta^{p} - 1$ is fulfilled for all $0 < p \le 1$.
\end{enumerate}
\vspace{2mm}
 These conditions are equivalent to Eq.~(\ref{eq:minselect-teq}). Consequently, the principle of minimum growth rate is derived theoretically.

\section{Summary and discussions}
\label{summary-discussion}
 To sum up, we have successfully given a theoretical explanation of the principle of minimum growth rate in many-body stochastic processes with multiplicative interactions. In the process of deriving the principle, we used the method of moment inequalities. This method made clear the correspondence between the solution of the transcendental equations and the divergence of moments as shown in Eq.~(\ref{eq:moment-inequalities}). When $s > 1$, the results coincide with those given by the method of Fourier transform. When $0 < s \le 1$, Eq.~(\ref{eq:gamma}) breaks down because of the divergence of the first-order moment. From the estimate of inequalities, we obtained the conditional inequality Eq.~(\ref{eq:gamma-condition}) which is replaced with Eq.~(\ref{eq:gamma}). 

 In order to obtain the principle of minimum growth rate, it is essential to estimate inequalities derived from the master equation of the processes. After performing Fourier transform, it is difficult to handle inequalities because of the existence of imaginary parts, which is probably the main reason why the principle had not been explained. It can be concluded that in the asymptotic analysis of this kind of stochastic processes, the method of moment inequalities are quite useful. Another application on weighted processes will be reported elsewhere\cite{FTOY2005}. 


\end{document}